\documentclass{aa}  
\usepackage{graphicx}
\usepackage{amsmath}
\usepackage{txfonts}
\begin{document} 

   \newcommand{\unit}[1]{\ensuremath{\, \mathrm{#1}}}
   \newcommand{\ramses}{{\sc ramses}}          
   \newcommand{\krome}{{\sc krome}}          
   \newcommand{\vapor}{{\sc vapor}}          
   \newcommand{\dispatch}{{\sc dispatch}}          
   \newcommand{\arepo}{{\sc arepo}}          
   \newcommand{\flash}{{\sc flash}}          
   \newcommand{\cloudy}{{\sc cloudy}}          
   \newcommand{\polaris}{{\sc polaris}}          
   \newcommand{\Fig}[1]{Fig.~\ref{fig:#1}}    
   \newcommand{\Figure}[1]{Figure~\ref{fig:#1}}    

   \title{The Bridge: a transient phenomenon of forming stellar multiples}

   \subtitle{Sequential formation of stellar companions in filaments around young protostars}

   \author{M. Kuffmeier
          \inst{1}\fnmsep\thanks{International Postdoctoral Fellow of Independent Research Fund Denmark (IRFD)}
          \and
          H. Calcutt\inst{2,3}
          \and
          L. E. Kristensen\inst{2}
          }

   \institute{Zentrum für Astronomie der Universität Heidelberg, Institut für Theoretische Astrophysik, Albert-Ueberle-Str. 2, 69120 Heidelberg
              \email{kuffmeier@uni-heidelberg.de}
         \and
             Centre for Star and Planet Formation (StarPlan) \& Niels Bohr Institute, University of Copenhagen, {\O}ster Voldgade 5--7, DK-1350 Copenhagen, Denmark
            \and Department of Space, Earth and Environment, Chalmers University of Technology, H\"{o}rsalsv\"{a}gen 11, 412 96, Gothenburg\\
             }

   \date{Received \today}

 
  \abstract
   {Observations with modern instruments such as \textit{Herschel} reveal that stars form clustered inside filamentary arms of $\sim$1 pc length embedded in Giant Molecular Clouds. On smaller scales ($\sim$1000 au), observations of, e.g., IRAS 16293--2422 show signs of filamentary `bridge' structures connecting young protostars in their birth environment.
    }
   {We aim to find the origin of bridges associated with deeply embedded protostars, by characterizing their connection to the filamentary structure present on GMC scales and to the formation of protostellar multiples. }
   {Using the magnetohydrodynamical code \ramses, we carry out zoom-in simulations of low-mass star formation starting from Giant-Molecular-Cloud-scales. We analyze the morphology and dynamics involved in the formation process of a triple system.}
   {
   Colliding flows of gas in the filamentary arms induce the formation of two protostellar companions at distances of $\sim$1000 au from the primary. After their birth, the stellar companions quickly ($\Delta t \sim 10$ kyr) approach and orbit the primary on eccentric orbits with separations of $\sim$100 au.
   The colliding flows induce transient structures lasting for up to a few 10 kyr connecting two forming protostellar objects that are kinematically quiescent along the line-of-sight. }
   {Colliding flows compress the gas and trigger the formation of stellar companions via turbulent fragmentation.
   Our results suggest that protostellar companions initially form with a wide separation of $\sim$1000 au. Smaller separations ($a\lesssim100$ au) are a consequence of subsequent migration and capturing.
   Associated with the formation phase of the companion, the turbulent environment induces the formation of arc- and bridge-like structures. 
   These bridges can become kinematically quiescent, when the velocity components of the colliding flows eliminate each other. However, the gas in bridges still contributes to stellar accretion later. 
   Our results demonstrate: bridge-like structures are a transient phenomenon of stellar multiple formation.
}


   \maketitle
%

\section{Introduction}
In the tradition of self-similar collapse \citep{Shu1977}, it has been common practice to model the formation of single stars from individual prestellar cores. For simplicity, 
cores are typically approximated as collapsing spheres \citep{Larson1969} detached from the environment. However, observations show that prestellar cores are part of larger-scale filaments threading the interstellar medium (ISM) \citep{Andre2010} causing deviations from spherical symmetry. In fact, stars form in different environments of Giant Molecular Clouds (GMCs) and evidence emerges that the majority of solar-mass stars form as part of multiple stellar systems \citep{Duquennoy-Mayor1991,Connelley2008,Raghavan2010}. In fact, recent surveys of Class 0 young stellar objects (YSOs; \citealt{Chen2013,Tobin2015}) reveal that multiples are already common in the early stages of star formation. However, the origin of multiples, and binaries in particular, is still debated. There are mainly two suggested mechanisms for binary formation, namely disk fragmentation \citep{Adams1989,Kratter2010} and turbulent fragmentation \citep{Padoan_turbfrag,Offner2010}.
It has been argued that the enhancement in separation to the closest neighbor of protostars at $\sim$100 au is caused by disk fragmentation,  while the companions at larger distances of $\sim$1000 au are either a sign of ejected companions or turbulent fragmentation. Determining the dominating mechanism is challenging though, given the computational costs involved in carrying out the necessary MHD simulations covering a large range of spatial scales. 

From an observational point, a well-studied example of a young binary system is IRAS 16293--2422 (hereafter IRAS 16293) \citep{Wotten1989,Mundy1992,Looney2000}. The projected distance between the two stars is 705 au \citep{Dzib2018} and both stars are connected via a small filamentary structure resembling a `bridge' between sources A and B \citep{Sadavoy2018,vanderWiel2019}.
Similar arc- and bridge-like structures have also been observed around other embedded sources such as IRAS 04191+1523 \citep{Lee2017}, SR24 \citep{Fernandez-Lopez2017} or L1521F \cite{Tokuda2014}.
Apart from that, polarization measurements around FUOri and in particular Z Cma reveal the presence of a stream extending several 100 au away from the central source \citep{Liu2016,Takami2018}.
Such structures are difficult to explain with the picture of an isolated, gravitationally collapsing, symmetrical core in mind. Therefore, models accounting for the protostellar environment provided by the GMC are required, such as has been done in recent `zoom-in' simulations \citep{Kuffmeier2017}. In these simulations, the starting point is a turbulent GMC, in which prestellar cores form consistently and where the formation process of stars and disks is studied by applying sufficient adaptive mesh refinement (AMR) around individual protostars. 
Based on such zoom-in simulations, \cite{Kuffmeier2018} illustrated the formation of a wide companion at a distance of approximately 1500 au from one of the investigated objects. 
In this paper, we focus our analysis on the gaseous filamentary structures present around this object, and we compare their morphology with observations of dense arc-like structures such as seen in, e.g., IRAS 16293. Furthermore, we investigate the formation process of two companions at distances of $\sim$1000 au that form due to compression inside filamentary arms within 90 kyr after the formation of the primary companion.  

The paper is divided into a brief description of the underlying method (Section 2), an analysis of the results (Section 3), a comparison of the results with observations (Section 4) and the conclusions (Section 5).


\section{Methods}
The simulations analyzed here are carried out with a modified version of the ideal MHD version of the adaptive mesh refinement (AMR) code \ramses\ \citep{Teyssier2002,Fromang2006}. 
We only give a brief summary of the `zoom-in' method here, and refer the reader to \cite{Kuffmeier2017} for a detailed description.
Our initial condition is a turbulent, magnetized GMC modeled as a cubic box of ($40$ pc)$^3$ in volume with periodic boundary conditions, and an average number density of 30 cm$^{-3}$ corresponding to about $10^5$ M$_{\rm \odot}$ of self-gravitating gas.
To circumvent computationally unfeasible time steps, we use sink particles as sub-grid models for the stars (for a description of the sink particle algorithm, please refer to \citet{Kuffmeier2016} and \citet{Haugboelle2018}).
Supernova explosions are used as drivers of the turbulence in the GMC, resulting in a velocity dispersion of the cold dense gas that is in agreement with Larson's velocity law \citep{Larson1981}. 

As a function for optically thin cooling, we use a table constructed by the computations of \citet{Gnedin_Hollon_2012}, who provide a publicly available Fortran code with corresponding database obtained by 75 million runs with \cloudy\ \citep{Ferland1998}, sampling a large range of conditions. 
The \cloudy\ simulations account for atomic cooling, but not for molecular cooling.
In principle, molecular cooling should be included for higher densities ($\rho \gtrsim 10^6 \unit{cm}^{-3}$ and $T<100 \unit{K}$), where it starts dominating over atomic cooling. Moreover, photoelectric heating is reduced for higher densities where UV radiation is attenuated.  
In contrary, cosmic rays as well as irradiation from individual (proto-)stars act as heating sources also for higher densities. 
To avoid extensive computational costs, we assume a simplified treatment in our models. To account for lower photoelectric heating due to UV shielding at higher densities, we taper down the temperature exponentially to $T=10 \unit{K}$ for number densities $n>200 \unit{cm}^{-3}$ \citep[see also][]{Padoan2016}. Protostellar heating is not accounted for in the model, and hence most of the gas in the densest regions is cold and quasi-isothermal.

In the first step (referred to as the parental run), the GMC is evolved for about 5 Myr and we apply a refinement of $16$ levels of $2$ ($l_{\rm ref}=16$) with respect to the length of the box $l_{\rm box}$, corresponding to a minimum cell size of $2^{-l_{\rm ref}} \times l_{\rm box} = 2^{-16} \times 40$ pc $\approx$ $126$ au. Several hundred sink particles form and evolve to different stellar masses during this run. 

In the next step, we rerun a simulation with higher resolution in the region and at the time of sink formation to better understand the individual accretion process of the selected sink. 
In other words, we `zoom-in' on the region of interest, which determines the name of the method.
We point out that we still account for the full domain of the GMC (i.e., the entire box of ($40$ pc)$^3$ in volume), when rerunning the simulation with higher resolution in the region of interest. 
Our follow-up illustrates the formation process of a triple stellar system for about 100 kyr after the formation of the primary star ($t=0$) modeled with a minimum cell size of 2 au until about $t=43$ kyr and a minimum cell size of 4 au thereafter. 
The secondary companion in this system forms after $t\approx 36$ kyr and the tertiary companion forms after $t \approx 74$ kyr. The accretion process of the primary sink (sink 4 in \cite{Kuffmeier2017}; sink b in \cite{Kuffmeier2018}) has already been previously analyzed until $t\approx 50$ kyr. 
In contrast to the previous simulations, 
we allowed maximum refinement for a larger region around the primary sink. To still be able to carry out the simulations for several $10$ kyr, we increased the density threshold for refinement of the highly refined cells and decreased the level of maximum refinement from 22 to 21, i.e., from minimum cell size of 2 au to minimum cell size of 4 au after $t=43$ kyr. In this way, we resolve the disk around the primary in less detail than in the previous studies. Compared to the previous models, we instead apply higher resolution for dense gas at distances $\sim 1000$ au from the primary. Therefore, we can simultaneously resolve the formation process of the companions together with the arc-structures associated with the primary more accurately as is the goal of this study.

In the following section, we present the morphology, formation and dynamics of the triple stellar system focusing in particular on the importance of gas streams associated with multiple star formation. 
We label the stars as primary A, first companion B and second companion C.

\section{Results}
\subsection{Filamentary structure throughout the scales}
\begin{figure*}
  \centering
  \includegraphics[width=0.8\textwidth]{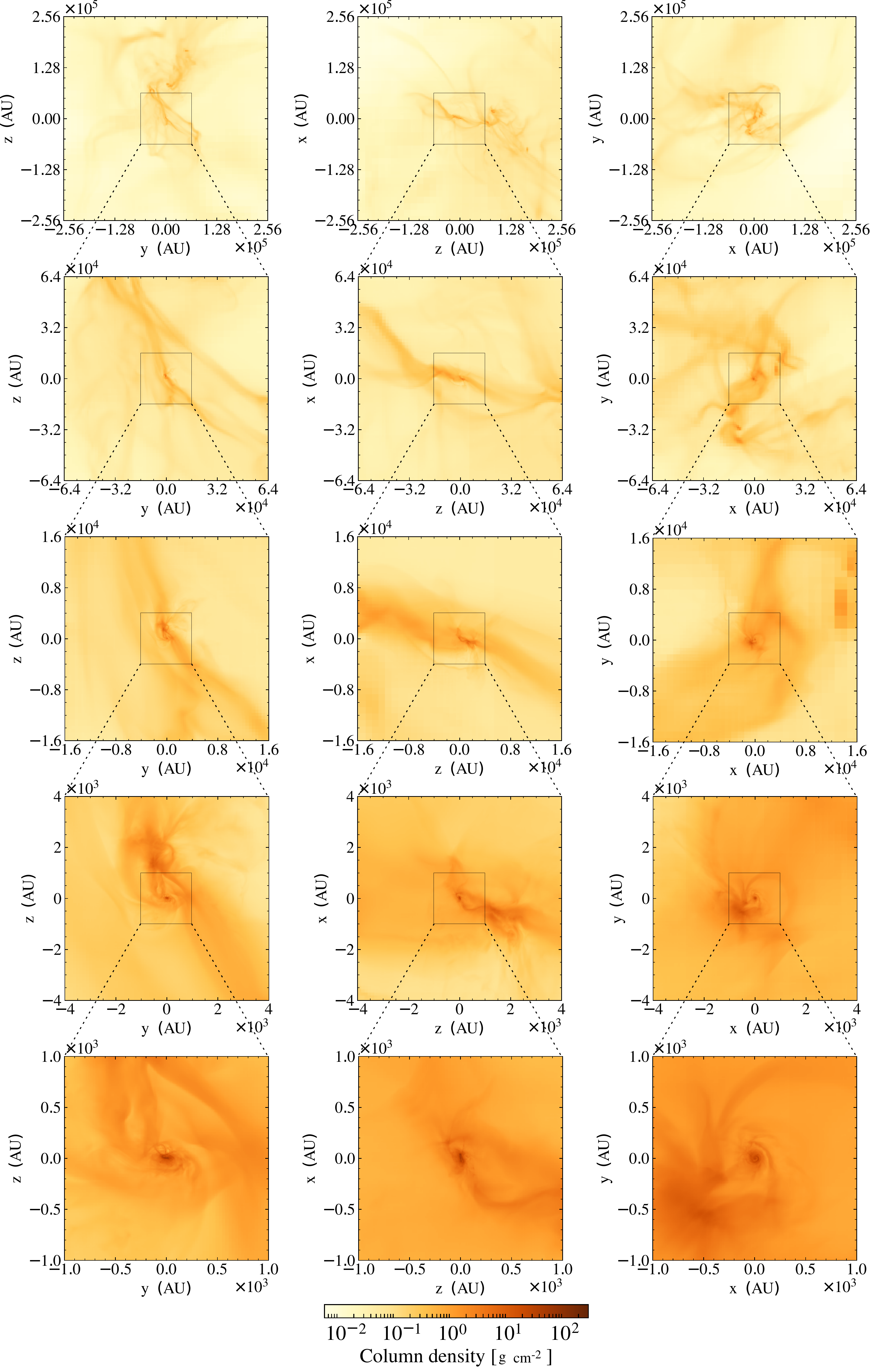}
  \caption{Illustration of the column density about $4$ kyr before the formation of the first companion B in the three planes of the coordinate system (left: $yz$-plane, middle: $zx$-plane, right: $xy$-plane) on different scales (row 1: $512 \times 10^3$ au $\approx 2.5$ pc, row 2: $128 \times 10^3$ au, row 3: $32 \times 10^3$ au, row 4: $8 \times 10^3$ au, row 5: $2 \times 10^3$ au). }
  \label{fig:prj_zoom}
\end{figure*}

\begin{figure*}
  \centering
  \includegraphics[width=0.48\textwidth]{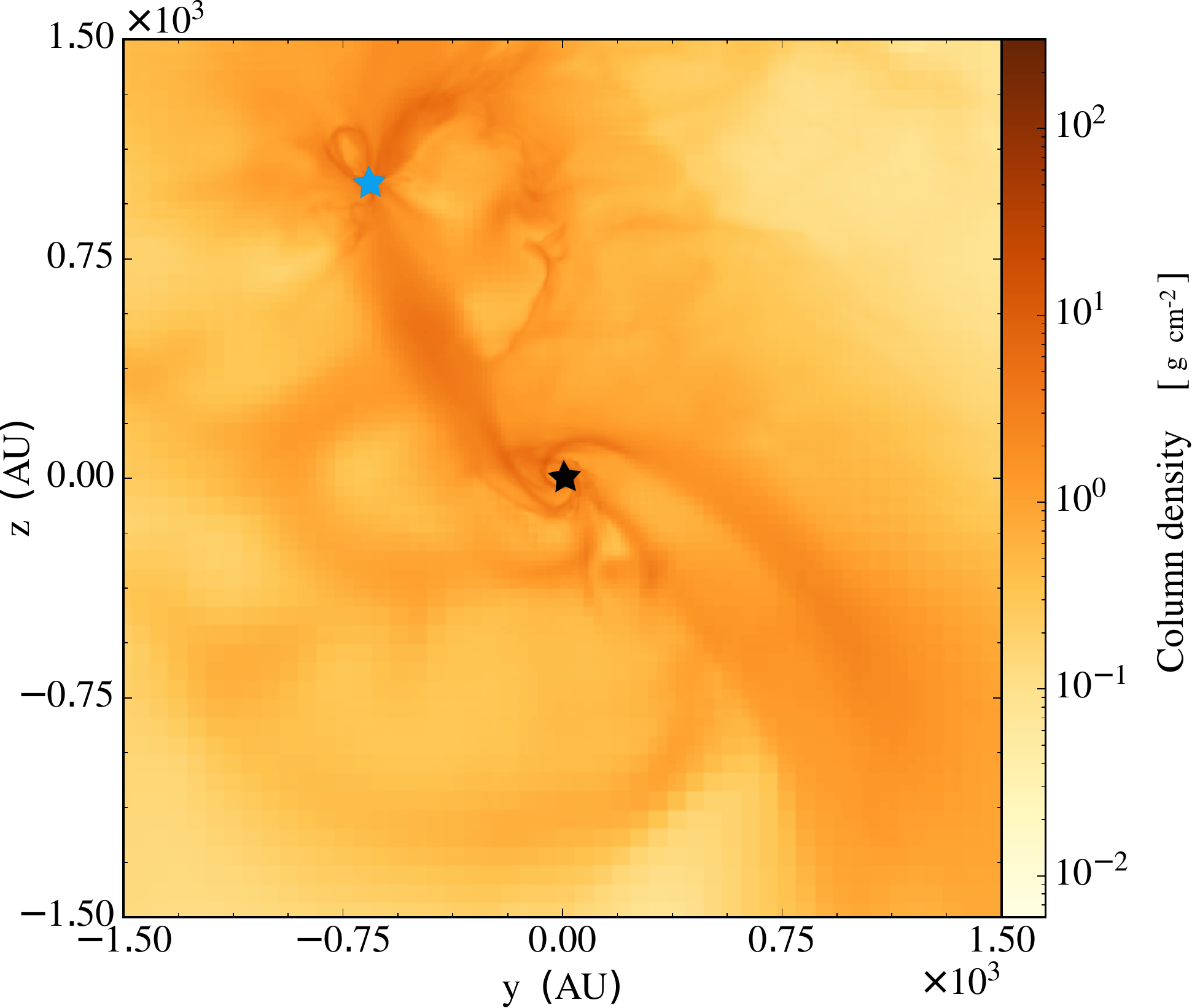}
  \includegraphics[width=0.48\textwidth]{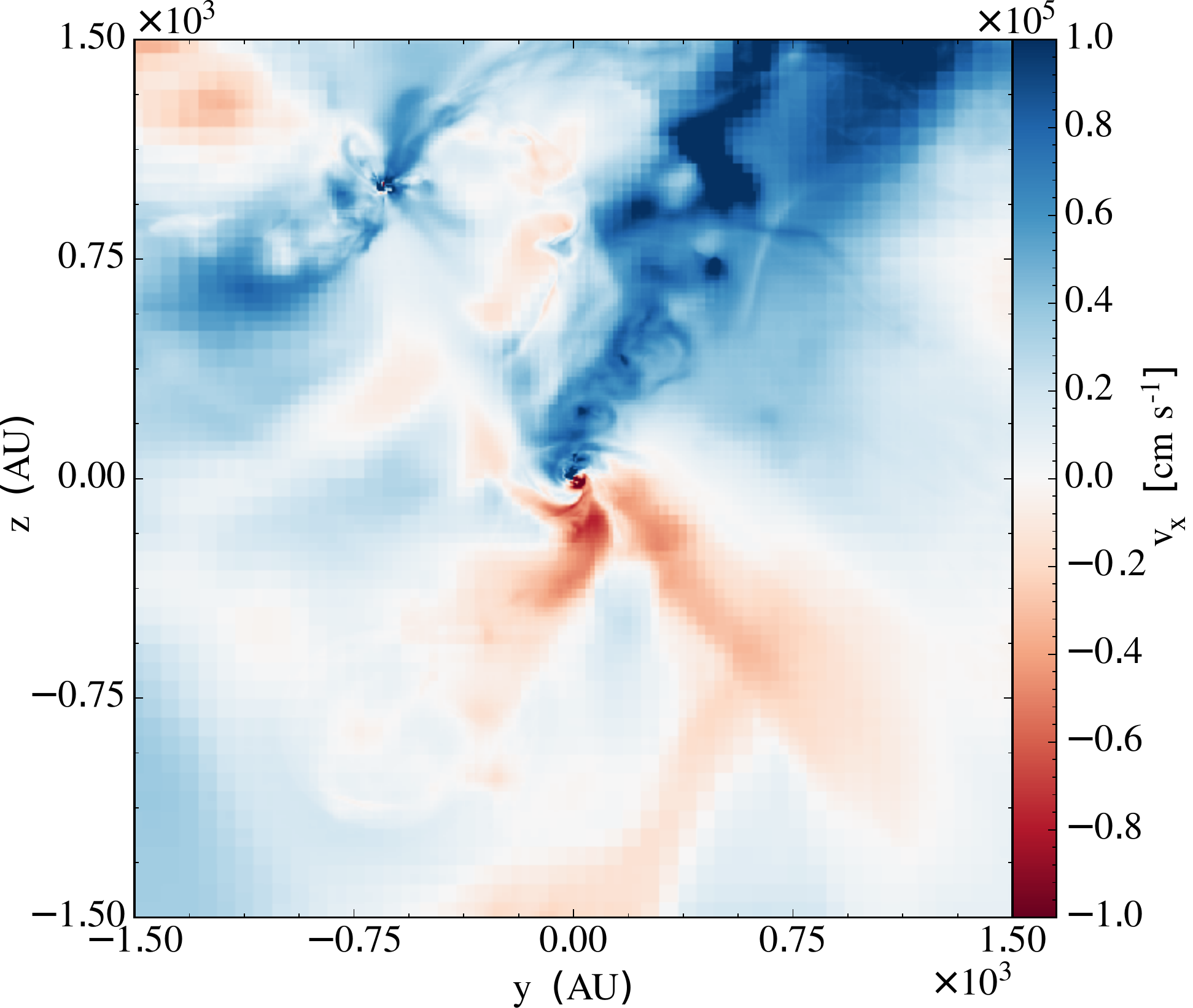}
  \caption{Illustration of the column density in the yz-plane (left panel) and density-weighted velocity along the x-axis relative to the systemic velocity of the young binary consisting of sink A and sink B (right panel) at time $t=t_0(B)+7 \unit{kyr} = 43$ kyr. The primary is located at the center and the displayed area is $ (3 \times 10^3 \unit{au})^2$. In the left panel, the black star symbol illustrates the location of sink A and the cyan star symbol shows the location of sink B.}
  \label{fig:prj_bridge1}
\end{figure*}

\begin{figure*}
  \centering
 \includegraphics[width=0.48\textwidth]{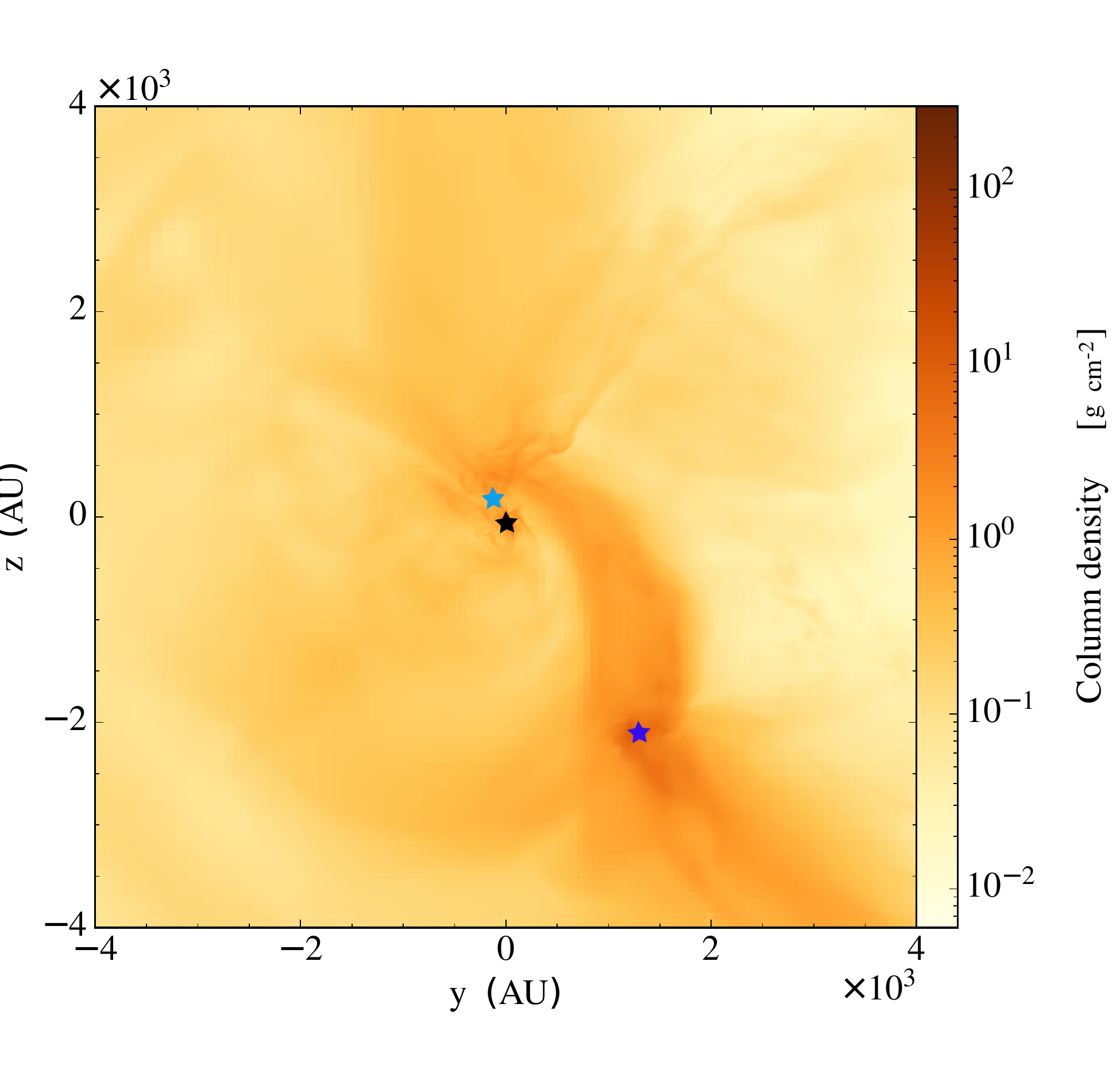}
  \includegraphics[width=0.48\textwidth]{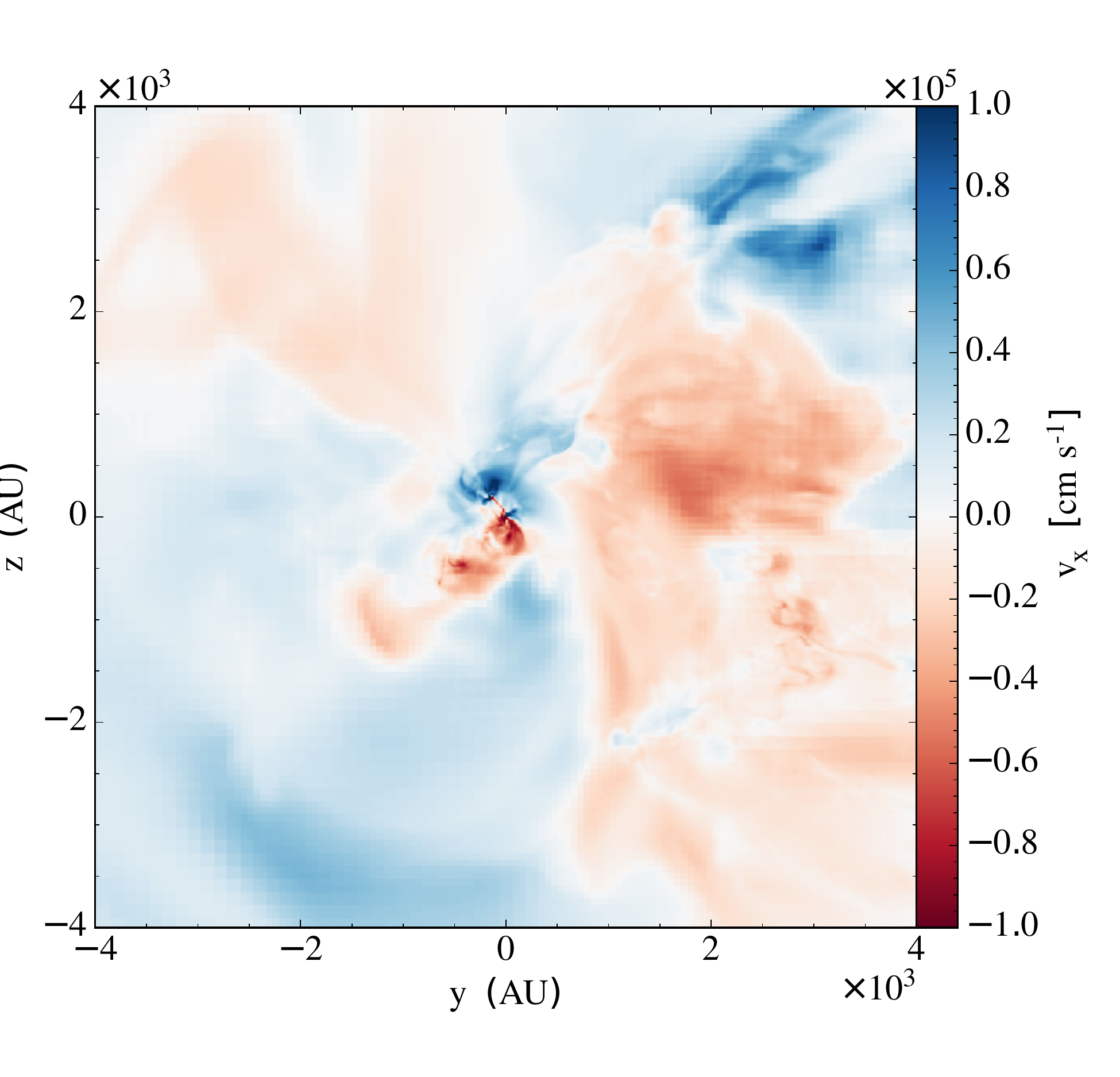}
    \caption{Illustration of the column density in the yz-plane (left panel) and density-weighted velocity along the x-axis relative to the systemic velocity of the binary consisting of sink A and sink B (right panel) at time $t=t_0(C)-4\unit{kyr}=70$ kyr. The primary is located at the center and the displayed area is $ (8 \times 10^3 \unit{au})^2$. In the left panel, the black star symbol shows the location of sink A, the cyan star symbol shows the location of sink B, and the blue star symbol shows the location, where sink C forms 4 kyr after this snapshot. }
  \label{fig:prj_bridge2}
\end{figure*}

To give a general overview of the environment in which the protostellar system is embedded during its formation, we show maps of the column density, $\Sigma$, in the three planes of the coordinate axes (\Fig{prj_zoom}). 
The maps are constructed in such a way that the primary A is at the center of the coordinate system and we illustrate $\Sigma$ at time $t=32 $ kyr $=t_0(B)-4$ kyr $=t_0(C)-42$ kyr.
At this point in time, the primary A has accreted to a mass of $M_{\rm A} \approx 0.29 \unit{M}_{\odot}$.
The panels on the left of \Fig{prj_zoom} show $\Sigma$ along the x-axis, the panels in the middle along the y-axis and the panels on the right along the z-axis. The plots in the top row cover an area of \begin{equation*}
    A_1=l_1^2=(5.12\times10^5 \unit{au})^2 \approx (2.5 \unit{pc})^2,
\end{equation*}
and the plots in the rows below have a length of $l_{i+1}=\frac{1}{4} l_i$ with respect to the preceding row, such that the fifth row covers an area of
\begin{equation*}
    A_5 = \left[ \left( \frac{1}{4} \right)^4 l_1 \right]^2 = [2000 \unit{au}]^2.
\end{equation*} 

The top row shows the presence of a filamentary arm of $\sim$ 1 pc in length in which the protostellar system is forming. 
Taking a closer look (row 2, especially along the z-axis), we see dense structures apart from the system of interest at projected distances of $\sim$0.1 pc that correspond to other forming or recently formed protostellar objects. 
We also see that the filament is more oriented along the z-axis than to the other two axes. 
When further zooming in on the region of interest (row 3), 
we see the dense elongated envelope around the primary A inside the filament. 
Examining the projections on scales of a few 1000 au (row 4) reveals the presence of a second dense region at a distance of about $\Delta r_{\rm AB} \approx 1500$ au from the primary star-disk system at the center. This accompanying accumulation of gas is the material from which the first companion B forms about 4 kyr later. 
The projections show the presence of several arms that are associated with the already formed primary A as well as with the forming companion B. 
Regarding the projections on the smallest scales around the protostar illustrates the morphology of the arms more clearly (row 5). 
Besides the connecting gas structure between the two objects, one can see the presence of dense arms feeding the young disk. The disk is rotationally supported at this stage up to a distance of $\approx 100$ au, where the azimuthally averaged rotational velocity $v_{\phi}$ drops to less than $0.8 v_{\rm K}$, where $v_{\rm K}=\sqrt{\frac{GM}{r}}$ is the Keplerian velocity \citep[see upper panel of Figure 13 in][]{Kuffmeier2017}.   
The (8000 au)$^2$ projection along the x-plane shows the presence of a gaseous arm extending to the lower right (row 4, left panel). 
In fact, companion C eventually forms at $\Delta r_{\rm AC}\approx 2100$ au about 43 kyr later inside this arm. 
The analysis above shows the ubiquity of filamentary structures on scales ranging from $\sim$1 pc down to $\sim$1000 au in \Fig{prj_zoom} during star formation. Stars preferentially form inside larger filaments consistent with observations of wide protostellar multiples \citep{Sadavoy2017}, and the arms present on smaller scales are important features of the heterogeneous star formation process.

\subsection{Formation of Quiescent Bridges}
In the left panel of \Fig{prj_bridge1} we show the column density in a region of 3000 au $\times$ 3000 au in the yz-plane at $t\approx 43\unit{kyr}$. 
The column density plot illustrates the presence of a bridge structure connecting sink A and sink B at this point in time. Briefly after the formation of sink B, the bridge-like structure emerges due to the compression of the filamentary arm seen in \Fig{prj_zoom}. During the approach of sink B to sink A, most of the mass inside the bridge region accretes onto sink A and sink B leading to a lifetime of the bridge-structure of about $\sim10$ kyr.
In the right panel of \Fig{prj_bridge2}, we show the velocity field with respect to the systemic velocity of sink A and sink B along the z-direction 
\begin{equation}
    \mathbf{v}_{\rm sys}(t) = \frac{ M_{\rm A}(t) \mathbf{v}_{\rm A}(t) + M_{\rm B}(t) \mathbf{v}_{\rm B}(t) }{M_{\rm A}(t) + M_{\rm B}(t)}
\end{equation}
with $M_{\rm A}$ ($M_{\rm B}$) and $\mathbf{v}_{\rm A}$ ($\mathbf{v}_{\rm B}$) being the mass and velocity of sink A (sink B). At this point in time ($t=43 \unit{kyr}$), the magnitude of the systemic velocity is $|\mathbf{v}_{\rm sys}| \approx -1.1\times 10^{4} \unit{cm}\unit{s}^{-1}$.
Comparing the column density with the density-weighted velocity structure perpendicular to the plane (line-of-sight-velocity) shows that the bridge structures have at most modest line-of-sight velocities ($v_{\rm x}<10^4\unit{cm}\unit{s}^{-1}$) with respect to the systemic velocity of sink A and sink B. That means that the bridge is kinematically quiescent along the line-of-sight at this point in time.     

In \Fig{prj_bridge2}, we show the same region as in the left panel of row 4 of \Fig{prj_zoom}, but $38$ kyr later (i.e., 4 kyr before the formation of the second companion C).
At this time, the primary A and companion B approach each other to form a binary system of the order of $100$ au in separation with masses of $M_{\rm A} \approx 0.49 \unit{M}_{\odot}$ and $M_{\rm B} \approx 0.25 \unit{M}_{\odot}$ (see subsection below). 
Compared to the earlier time, the relatively broad gaseous arm (lower right in the yz-plane, left in the zx-plane and upper part in the xy-plane of row 4 in \Fig{prj_zoom}) is denser and more pronounced due to compression. 
At this point in time, the projection along the x-axis again shows a bridge-like structure connecting the central binary system with the forming additional companion C.
In general, the turbulent motions inherited from the GMC induce a rather complex velocity structure (in particular in the yz- and xy-planes). 
Following the dynamics of the system from $t=t_0[B]-4$ kyr snapshot until the formation of companion C, we see that the left part of the fork-like structure visible at the bottom right in the yz-projection in row 2 and 3 of \Fig{prj_zoom} merges with the longer arm. 
This compression contributes to the accumulation of mass in the arm that eventually leads to the formation of companion C. 

Similar to the bridge shown in \Fig{prj_bridge1}, it is also evident that the bridge shown in \Fig{prj_bridge2} is a result of the larger filamentary structure presented in \Fig{prj_zoom}. Looking at the line-of-sight velocity ($v_{\rm x}$) relative to the systemic velocity (right panel in \Fig{prj_bridge2}) shows the variations of the velocity field in the surroundings. Although the velocity in the bridge has a mildly negative line-of-sight velocity ($v_{\rm x} \sim -10^4 \unit{cm}{s}^{-1}$), the plot nevertheless shows a transition from positive to negative velocities associated with the bended arm. The plot shows that bridges become kinematically quiescent, once the flows with different orientation cancel out each other. 
In general, the dynamical history and evolution of the triple system demonstrate that bridge-like structures occur as a side-effect during the formation of multiple star systems.

\begin{figure}
  \centering
  \includegraphics[width=0.5\textwidth]{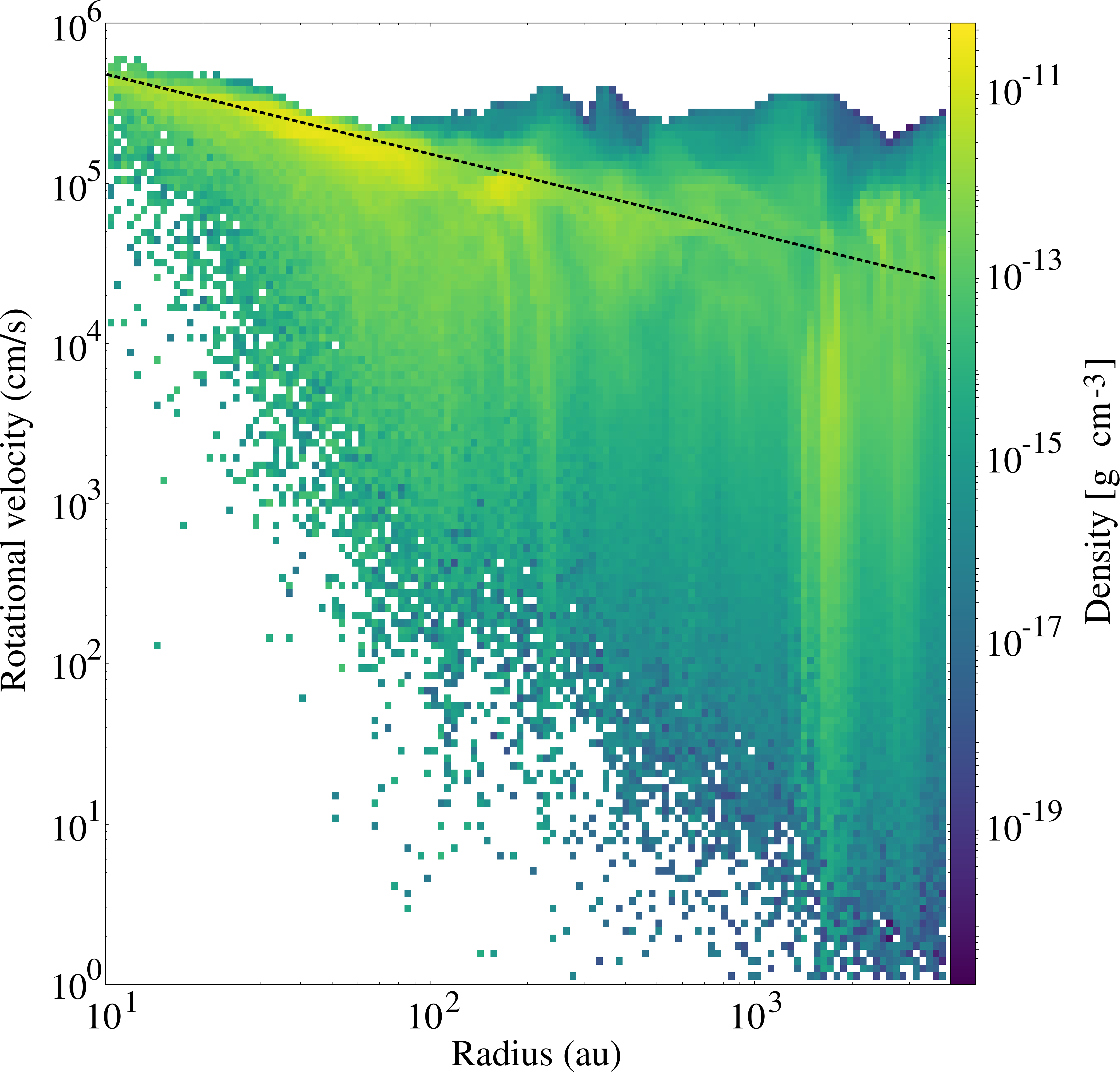}
  \caption{Phase diagram illustrating cylindrically azimuthal velocity $v_{\phi}$ in a cubical region of ($8 \times 10^3$ au)$^3$ around the primary about $4 \times 10^3$ yr before formation of companion B. The rotational axis is chosen as the orientation of the angular momentum vector computed for a sphere around the primary A of radius 1000 au at this point in time. }
  \label{fig:r_v_rho}
\end{figure}

\subsection{Velocity Structure}
In this section, we present the velocity field around primary A during the early evolution shortly before the formation of companion B. 
We plot the magnitude of the rotational velocity gas $v_{\phi}$ for all cells within a radial distance of $r=4000$ au from the primary at $t =32$ kyr, i.e., $t=t_0[B]-4$ kyr (\Fig{r_v_rho}). 
The color is used to display the density of each cell and the black dashed line shows the Keplerian velocity 
\begin{equation}
    v_{\rm K} = \sqrt{ \frac{GM}{r} }
\end{equation}
at this point in time for the sink mass of $M_{\rm A}\approx0.29$ M$_{\odot}$, where $G$ is the gravitational constant and $r$ is the radial distance from the sink.
At relatively small distances from the primary ($r \lesssim 100$ au), the plot shows the approximately Keplerian profile ($v_{\phi}\propto r^{-0.5}$) of the dense gas associated with the rotationally-supported disk.   
Cells with large deviations from the Keplerian profile at $r \lesssim 100$ au have low densities and are not located in the midplane of the young disk.
The disk truncates at a radius of $r \sim 100$ au as seen by the drop in density in the plot. 
However, looking carefully at the diagram one can see some cells at a distance of $r\approx 150$ au to $200$ au of relatively enhanced density $\rho > 10^{-12}$ g cm$^{-3}$ and velocity magnitude $v\approx 10^5$ cm s$^{-1}$.
In fact, this small characteristic is caused by the small gas stream visible in the lower right panel in \Fig{prj_zoom}.     
The velocity profile scaling slightly steeper than the Keplerian relation $v \propto r^{-0.5}$ is consistent with a gas parcel spiralling toward the central protostar.

Apart from that feature, the densities generally drop with increasing distance up to $r \approx 1000$ au,
where the gas accumulates to form the companion. 
In particular, one can see a wide spread in velocity magnitude ($10^3 \unit{cm} \unit{s}^{-1} \lesssim v \lesssim 10^5 \unit{cm} \unit{s}^{-1}$) of the dense gas associated with the formation of companion B.  
Accounting also for the gas at lower densities $\rho \lesssim 10^{-15}$ g cm$^{-3}$ at $\approx 3\times 10^3$ au 
shows an even larger spread of more than three orders of magnitude in velocity magnitude ($3\times 10^2 \unit{cm} \unit{s}^{-1} \lesssim v \lesssim 7\times 10^5 \unit{cm} \unit{s}^{-1}$). 

\begin{figure}
  \centering
  \includegraphics[width=\columnwidth]{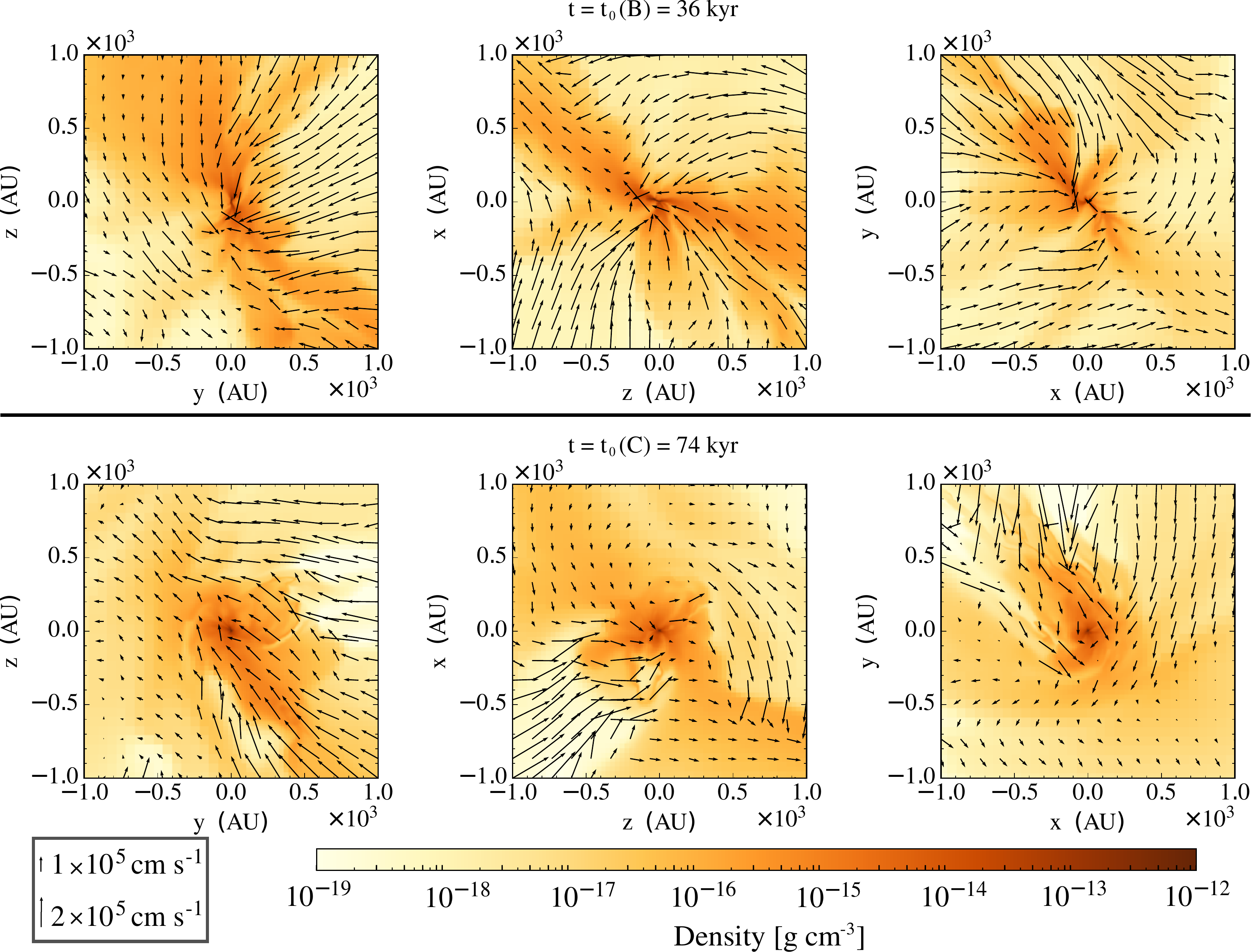}
  \caption{Illustration of the density distribution at the time $t = t_{\rm B,0}$ of the formation of the first companion B (upper panels) and the formation of the second companion C at $t = t_{\rm C,0}$ (lower panel). The panels show slices of the three planes spanned by the coordinate system (left: $yz$-plane, middle: $zx$-plane and right: $xy$-plane) with the position of the forming sink at the center. The arrows show the velocity with respect to the systemic velocity in the corresponding plane for every 50th data point in the plane. The length of the arrows scales linearly with the velocity magnitude. In the lower left corner, the length corresponding to $10^5$ cm s$^{-1}$ and $2\times10^5$ cm s$^{-1}$ is shown.}
  \label{fig:comp1+2_slc}
\end{figure}

Analyzing the structure of the velocity field also shows the diversity of the orientation of the vector field.
In \Fig{comp1+2_slc} we illustrate the velocity orientation around companion B and companion C at the time of their formation in detail with respect to the systemic velocity. 
We plot the density distribution and velocity vectors around companion B (upper panels) and companion C (lower panels) in slices ($2000$ au)$^2$ of the three planes spanned by the coordinate axes (left: yz-plane, middle: zx-plane and right: yz-plane). 
The plots clearly show the different orientation of the velocity vectors leading to the compression that eventually causes the formation of the individual companions. Moreover, the velocity field in the xy-plane shows that the binary system of sink A and sink B moves toward the forming companion, thereby eventualy sweeping up part of the material in the bridge at later times.

In the following subsection, we analyze the formation process of the companion explicitly.
We interpret the differences in velocities together with the abundance of filamentary structures as a consequence of the underlying turbulence present in the GMC cascading down to smaller scales. 

\begin{figure}
  \centering
  \includegraphics[width=\columnwidth]{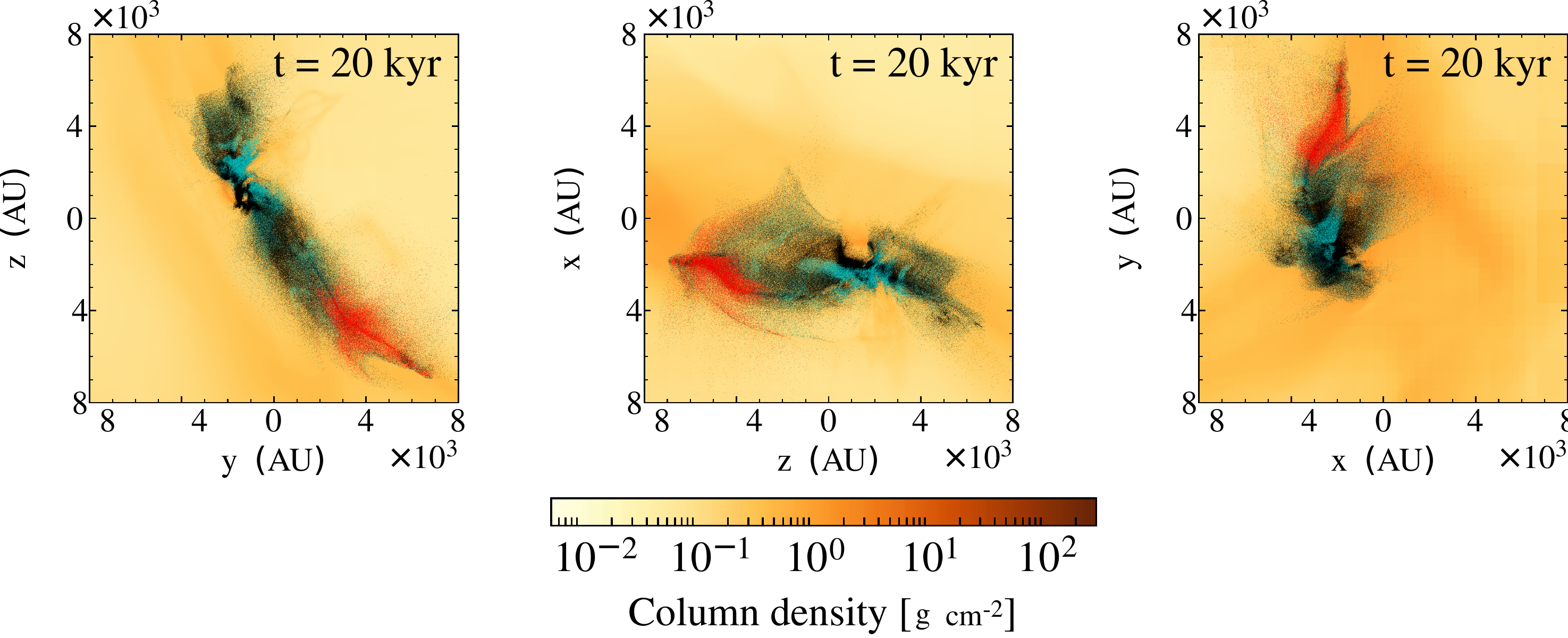}
  \caption{Illustration of the column density in the three planes of the coordinate system (width: $1.6 \times 10^4$ au; left: $yz$-plane, middle: $zx$-plane, right: $xy$-plane) at time $t=20 \times 10^3$ yr after formation of the primary A. The colored dots illustrate the origin and dynamics of accreting gas of the individual sinks. Black (cyan, red) dots represent particles that are located within a distance of 30 au from the primary A (B, C) at $t=90\times 10^3$ yr.  }
  \label{fig:prj_part}
\end{figure}

\begin{figure}
  \centering
  \includegraphics[width=\columnwidth]{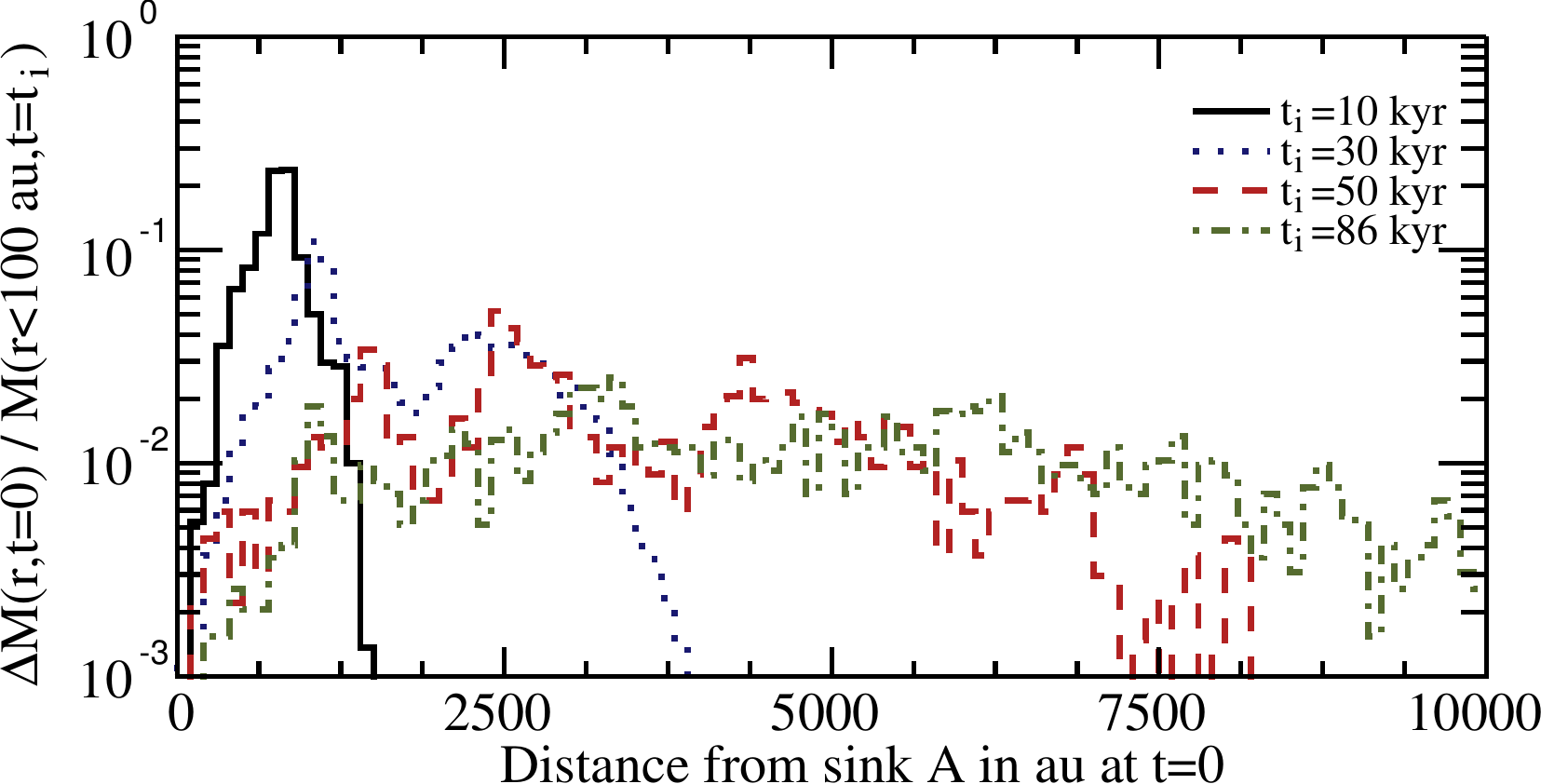}
  \includegraphics[width=\columnwidth]{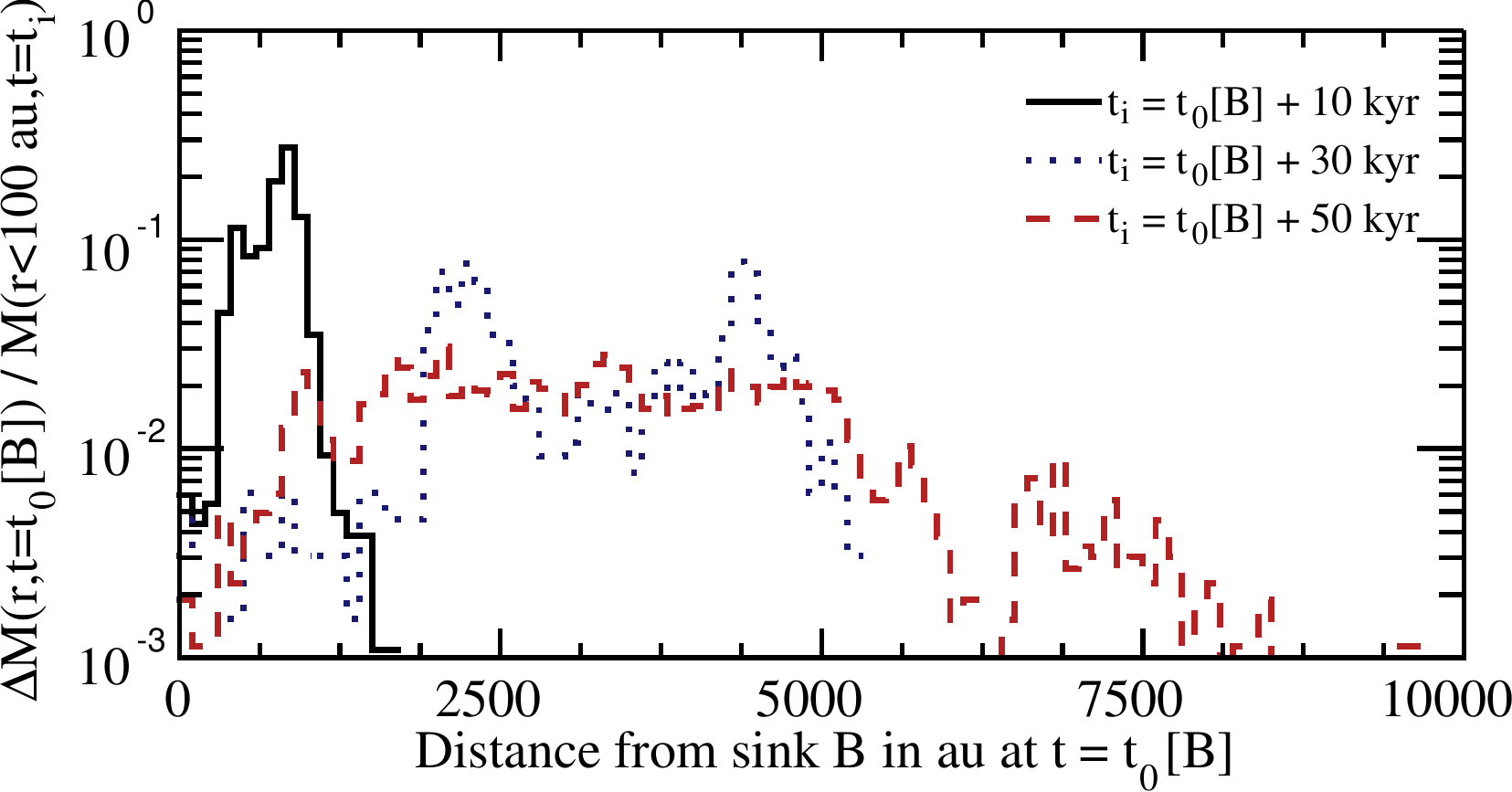}
  \caption{Origin of gas for sink A and sink B at $t=10$ kyr (black solid line), $t=30$ kyr (blue dotted line), 50 kyr (red dashed line) and $t=86$ kyr (green dashed-dotted line) after the formation of the individual sinks. }
  \label{fig:origin_hist}
\end{figure}

\begin{figure}
  \centering
  \includegraphics[width=\columnwidth]{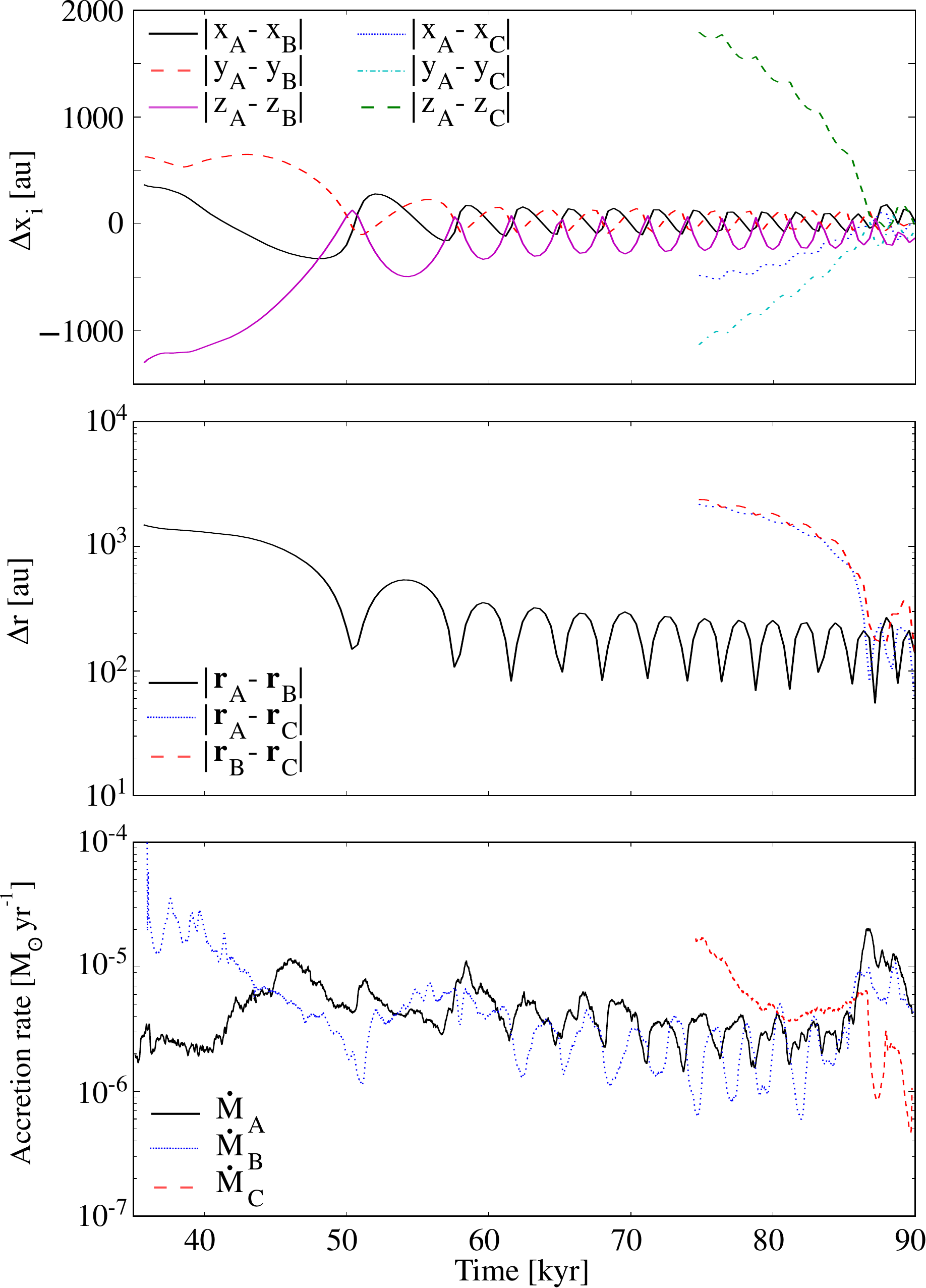}
  \caption{Evolution of the distance between the different objects of the multiple stellar system.
  The upper panel shows the difference between sink A and B in $x$ (black solid line), $y$ (red dashed line) and $z$ (magenta solid line), as well as the difference between sink A and C in $x$ (blue dotted line), $y$ (cyan dash-dotted line) and $z$ (green dashed line). 
  The middle panel shows the absolute distance $r$ between sink A and B (black solid line), sink A and C (blue dotted line) and sink B and C (red dashed line).
  The lower panel shows the accretion profile for the three sinks involved from $t=35 \times 10^3$ yr to $t=90 \times 10^3$ yr after formation of the primary. The black solid line represents the primary A, the blue dotted line corresponds to companion B and the red dashed line corresponds to companion C.}
  \label{fig:dist_sinks}
\end{figure}


\begin{figure}
  \centering
  \includegraphics[width=\columnwidth]{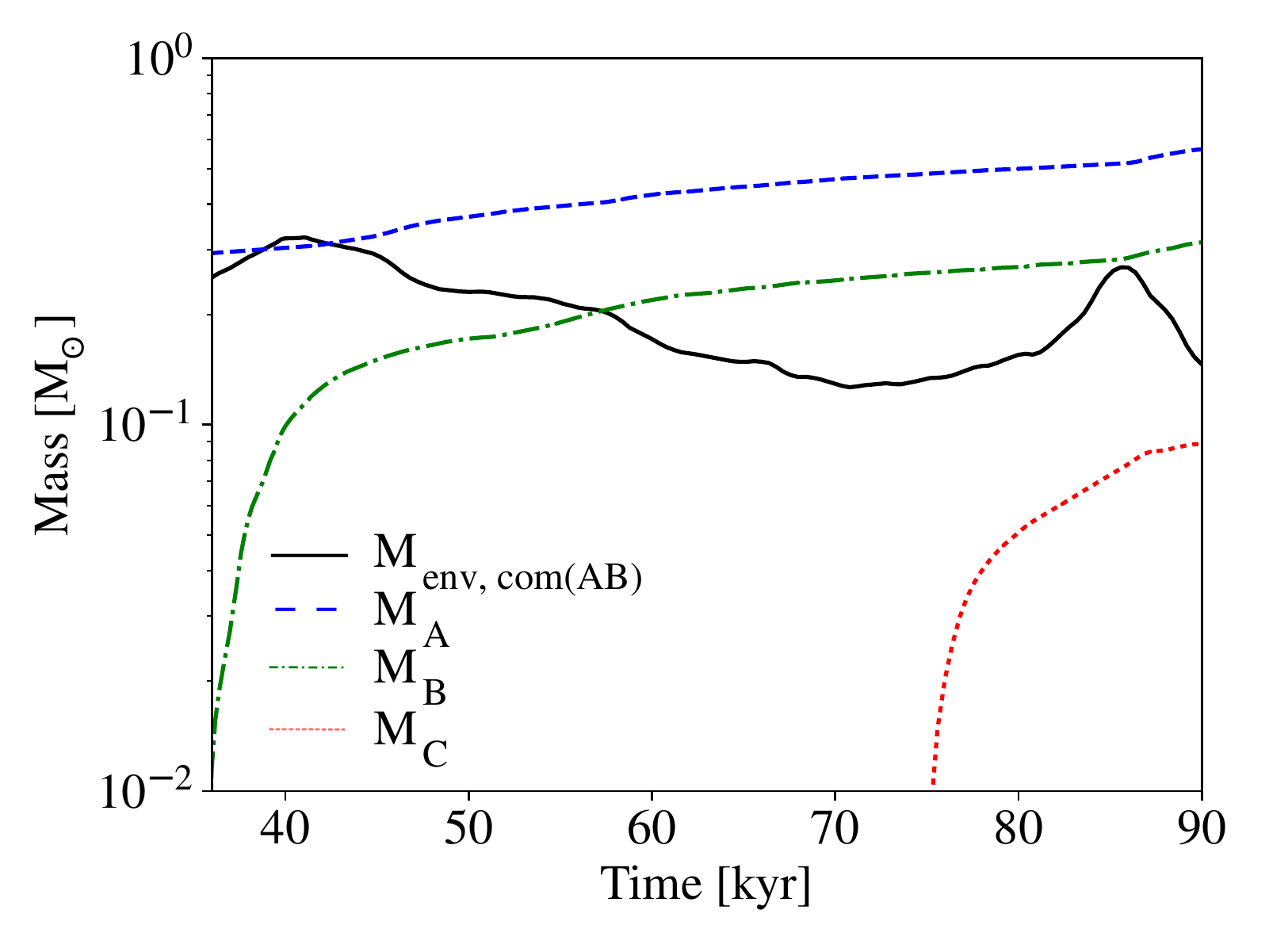}
  \caption{The plot shows the evolution of mass located within a distance of 1000 au from the center of mass of the primary and secondary (black solid line), mass of sink A (blue dashed line), mass of sink B (green dashed-dotted line) and mass of sink C (red dotted line).}
  \label{fig:Menv}
\end{figure}

\subsection{Formation of companions}
The critical radius of an isothermal sphere supported by gas pressure against gravitational collapse, i.e., a Bonnor-Ebert sphere, is given by 
\begin{equation}
R_{\rm BE} = 0.49 \frac{c_s}{\sqrt{G \rho_0}},
\end{equation}
where $c_s$ is the sound speed, G is the gravitational constant and $\rho_o$ is the outer density \citep{Ebert1955,Bonnor1956}.
As a convenient estimate of $R_{\rm BE}$ in practical units, one can use 
\begin{equation}
\left[\frac{R_{\rm BE}}{\unit{pc}} \right]= 1.88 \left[\frac{T}{\unit{K}} \right]^{0.5} \left[\frac{n}{\unit{cm}^{-3}}\right]^{-0.5},
\end{equation}
which would yield a radius of about $10^4$ au assuming a number density of $10^4\unit{cm}^{-3}<n<10^5\unit{cm}^{-3}$ and temperature $T=10 \unit{K}$ considered as typical for back-of-the-envelope calculations of the collapse of a solar mass star. 
It is evident that the formation of both companions deviates from such a classical collapse scenario for single stars, as also indicated by the relatively small collapsing region of only $\sim$100 au just at the location where the individual companions form. 
Instead, the companions form rather as a consequence of turbulent fragmentation inside the elongated heavily perturbed prestellar core similar to what has seen in dedicated core collapse simulations with turbulence \citep[e.g.][]{Seifried2013}.

In the following, we investigate the formation of the first companion in more detail. 
Shortly after formation of the primary A, gas predominantly approaches the sink from within the filament resulting in non-isotropic accretion. 
Given that the inflowing gas has angular momentum with respect to the star, not all of the gas in the flow accretes onto the protostar. 
Instead, part of the gas is deflected by the gravitational field of the protostar and passes by the protostar. 
However, gas also approaches the protostar from the opposite direction and hence compresses the gas to form companion B at a distance of $\approx1500$ au from the primary (see accumulation of gas in the projection plots in \Fig{prj_zoom} and in the slice plots, upper panel \Fig{comp1+2_slc}).

Following the system further in time, the two stars approach and orbit each other eccentrically with a separation between $\lesssim 100$ au and $\sim$300 au. 
While this happens, gas also passes by the primary star and gets compressed in a dense arm similar to the scenario before the formation of the first companion (see \Fig{prj_bridge2}).
As a consequence of this, the second wide companion C forms at a distance of about 2100 au from the close binary system.  

To give a better overview of the evolution of the gas contributing to the formation process of the three different stars, we show maps of the column density of size ($1.6 \times 10^4$ au)$^2$ along the three coordinate axes for four different times ($t=20$ kyr, $t=50$ kyr, $t=70$ kyr and $t=90$ kyr after formation of the primary) in \Fig{prj_part}. 
The maps are centered at the location of the primary and the dots in the plot represent gas that is located within 30 au at $t=90$ kyr from the primary A (black dots), secondary B (blue dots) and tertiary C (red dots). 
Using tracer particles, we can constrain the origin of the accreting gas for the individual sinks. 
The figure clearly illustrates that most of the material accreting onto the triple system is indeed located in the dense filamentary arm. 

In \Fig{origin_hist}, we plot, how far away the gas that is located within 100 au from sink A (upper panel) and sink B (lower panel) at $\Delta t=10(30,50,86)$ kyr after sink formation ($t_{\rm form}$), was located at $t_{\rm form}$. 
The plot demonstrates that both sinks initially accrete the collapsing gas in their vicinity of $\sim$ 1000 au. 
However, at later times a significant fraction of the mass stems from distances initially several 1000 au from the sinks. 
Gas accreting from distances beyond the scale of a Bonnor-Ebert sphere is inconsistent with the expected accretion pattern of a traditional core collapse models. 
However, consistent with observations, the sinks form in elongated filaments. 
As shown in \Fig{prj_part} and \Fig{origin_hist}, accretion inside these filamentary birth environments allows stars to accrete gas from initially far distances.

Moreover, \Fig{prj_part} shows that all of the three objects share the same reservoir
although the reservoir of companion C is a bit more distinct. 
This is not surprising, considering that companion C is the youngest and least massive of the three objects.
Furthermore, as illustrated in \Fig{dist_sinks}, sink B approaches sink A during the evolution and the two sinks accrete gas as a binary system of smaller separation for some time before the formation of companion C at a larger distance from the---by then---relatively close binary system of separation $\sim$100 au. 
The sinks initially have the largest separation in $z$-direction (magenta solid line for $\Delta z_{\rm AB}$ and green dashed line $\Delta z_{\rm AC}$ in \Fig{dist_sinks}), which reflects the fact that the filamentary arm is predominantly oriented along the $z$-axis. 
The separation between both companions to the primary is initially largest before the sinks approach each other. In particular, companion C and the binary star consisting of A and B approach each other faster than companion B approaches A after its formation
due to the stronger gravitational interactions between the sinks.
At the time of formation of companion C $t_{\rm 0,C}$, the mass of primary A ($M_{\rm A}(t_{\rm 0,C}) \approx 0.49 \unit{M}_{\odot}$) together with the additional mass of companion B ($M_{\rm B}(t_{\rm 0,C}) \approx 0.26 \unit{M}_{\odot}$) in the vicinity of A corresponds to a higher gravitational potential than at the earlier time of formation of companion B $t_{\rm 0,B}$, where the primary had a mass of $M_{\rm A}(t_{\rm 0,B}) \approx 0.29 \unit{M}_{\odot}$. 

\subsection{Accretion and evolution of the protostellar multiple}
Investigating the accretion profile of the different sinks (lower panel of \Fig{dist_sinks}), 
we see a direct effect of the dynamics on the accretion process of the sinks. 
Focusing on the profile of companion A and B first, the accretion rate of the primary increases when companion B comes closer. 
Later, the eccentric orbits of companion B around A cause a periodic pattern in the accretion rates of both primary A and companion B. 
A similar effect is also evident when the second companion approaches the binary system consisting of A and B.
To understand the accretion process more clearly, we plot the evolution of mass that is enclosed within a radius of 1000 au from the center of mass of the primary and the secondary
\begin{equation}
\mathbf{r_{\rm com}} = \frac{ m_{\rm A} \cdot \mathbf{r_{\rm A}} + m_{\rm B} \cdot \mathbf{r_{\rm B}}}{m_{\rm A} + m_{\rm B}},  
\end{equation}
where $m_A$ ($m_B$) represents the mass of the primary A (secondary B) and $\mathbf{r}_A$ ($\mathbf{r}_B$) corresponds to the position of the primary A (secondary B) in \Fig{Menv}.
The plot shows an increase in enclosed mass around the binary system for the approach of companion C seen in \Fig{dist_sinks}.   
Hence, the mass reservoir for accretion onto the binary system of sink A and B is refueled leading to the increase in accretion rate seen in the lower panel of \Fig{dist_sinks}.
In contrast the less massive approaching sink now has to share its mass reservoir with the already established binary system, 
and hence its accretion rate drops. 
Quantitatively, when the tertiary star approaches the system to a distance of about 200 au, its accretion rate from about $\dot{M} \approx 6\times 10^{-6} \unit{M}_{\odot} \unit{yr}^{-1}$ to less than $10^{-6} \unit{M}_{\odot} \unit{yr}^{-1}$ within less than 1 kyr, 
while the accretion rates of companion B and especially primary A increase by up to a factor of 10 within only a few kyr. 

Our results show the importance of dynamical interactions on the accretion process of young deeply embedded protostars. 
Without the presence of gas during this stage, the migration process of the companion(s) would be free of dissipation due to the lack of accretion.
Consequently, the secondary B would approach and leave the primary again on a hyperbolic trajectory.      
However, the two stars are still deeply embedded. The surrounding gas has a dissipative effect on the secondary through gas accretion during migration toward the primary. 
Using smoothed particle hydrodynamics (SPH) simulations, \cite{BateBonnell1997} carried out a parameter study for accreting circular binary systems with constant infalling specific angular momentum demonstrating that the separation of binaries decreases, even if the specific angular momentum of the infalling gas is much larger than the specific angular momentum of the binary.

The focus of this work is the morphology of the bridge structure during stellar multiple formation, and an in-depth analysis of the evolution of binary separation and angular momentum transfer is beyond the scope of this paper. 
However, consistent with previous models \citep{BateBonnell1997,Offner2010}, our results suggest a characteristic sequence for the formation process of multiple stellar systems:
\begin{enumerate}
\item formation of a primary as a consequence of gravitational collapse in a deformed prestellar core; 
\item formation of secondary in the filamentary arm connected to the primary due to contraction of mass at distances of $\gtrsim 1000$ au induced by the underlying turbulence in the GMC, consistent with an observed peak at $\sim$3000 au for YSO Class 0 objects in Perseus \citep{Tobin2016_multisurvey}; 
\item migration of the secondary toward the primary induced by the gravitational potential of the relatively massive primary; 
\item due to the interaction of gravity and accretion, the secondary is captured by the primary and forms an eccentric binary system with characteristic separation of $\sim$ 100 au consistent with the observed peak in the distribution of protostellar separation for Class 0 and even more for Class I objects \citep{Tobin2016_multisurvey}.
\end{enumerate}
Considering subsequent components, our models suggest the same initial sequence as for the secondary (steps 2 to 4). 
However, different to the two-body scenario, tidal interactions in three-body system also imply dissipation that can possibly lead to capturing of companions even without any gas. Possibly, one of the components is ejected during the interaction potentially leading to the formation of binary systems with smaller separation. 

\section{Discussion}

\subsection{Constraining the origin of protostellar companions}
There are two suggested mechanisms for the formation of stellar multiples: disk fragmentation \citep{Adams1989,Kratter2010} and turbulent fragmentation \citep{Padoan1995,Padoan_turbfrag,Offner2010}.
Per definition, the former can only occur in protostellar disks, i.e., on scales of $\lesssim 100$ au, while turbulent fragmentation is predominantly acting on larger scales of $\gtrsim 1000$ au.  
Although statistical constraints of main-sequence stellar binaries and multiplicities have been known for decades \citep{Duquennoy-Mayor1991},
only recently has it become feasible to constrain multiplicity during the protostellar phase.
Using the Karl G. Jansky Very Large Array (VLA), \cite{Tobin2016_multisurvey} provide constraints on the  multiplicity fraction during the protostellar phase for Class 0 and Class I YSOs in Perseus.
The survey shows a bimodal distribution for the protostellar separation in the Class 0 with a peak at $\sim$75 au and another peak at $\sim$3000 au. 
The authors attribute the inner peak to disk fragmentation and the outer peak to turbulent fragmentation though they also acknowledge that the lower number of binaries with separation of $\gtrsim 1000$ au for Class I might be a consequence of inward migration of companions formed by turbulent fragmentation.  
To properly constrain the formation mechanism, 
computationally expensive models accounting for the turbulence in the ISM are necessary.

The selected primary forms as a consequence of gravitational collapse of dense gas within a perturbed core structure. In contrast, the formation of the companions occurs inside the gaseous arms that are connected to the primary in a different manner. Tracing the evolution at the location close to sink formation, we see for both companions that their formation may be understood as a consequence of colliding flows. The gas inside the long filamentary arm feeds the star, while the velocity field around it has a different orientation, and hence compresses the gas enough to cause sink formation. One may wonder whether the sink only forms because of insufficient resolution of the angular momentum present in the flow structure at $\approx 2$ au resolution. To test the robustness of sink formation, we conducted comparison runs with $l_{\rm ref}=23$(24,25,26,27) corresponding to minimum cell sizes of $\approx1$ ($\approx0.5$ au, $\approx0.23$ au, $\approx0.123$ au, $\approx0.061$ au) as shown in the appendix. We confirm the formation of the sink in all of these comparison runs demonstrating the robustness of companion formation.

\subsection{Dynamical evolution of the protostellar companions}
Recently, \cite{Munoz2019} thoroughly carried out 2D hydrodynamical simulations with the moving mesh code \arepo\ \citep{Springel2010} of an accreting equal-mass binary. In contrast to our results, they find an increase in stellar separation $a$ rather than a decrease; the increase in separation is $\approx 5 \times$ stronger for the circular case than for the eccentric case $e=0.6$.
However, their setup is quite different from our zoom-in setup. In our simulations the companion forms in its turbulent birth environment, where it is initially unbound, and gets captured by the primary at a later stage of evolution.
In contrast, their simulations start with a binary star that is already in a bound state, and which evolves for many more orbits ($N_{\rm orbit}=3500$) in an idealized 2D setup. Moreover, our results account for the effects of magnetic fields that can transfer angular momentum away from the gas close to the star. Therefore, it is difficult to directly compare our results of a young forming protostellar binary with the longer term evolution of an already existing binary system in a less violent environment.  
Another significant difference between our scenario and a scenario of an already established binary system is the change in mass ratio of the binary components. As \cite{Munoz2019} pointed out the mass ratio in their setup is $q=1$, whereas in our scenario the ratio varies and quickly increases briefly after the formation of companion B.

For a conceptual understanding of the effects of mass ratio $q$ and mass accretion rate of the binary $\dot{M}_{\rm b}$, we discuss the fiducial case of an accreting circular binary as analyzed in detail by \cite[e.g.,][]{BateBonnell1997}. 
Taking the time derivative of the angular momentum around its centre of mass 
\begin{equation}
    L_b = \sqrt{G M_{\rm b}^3 a} \frac{q}{(1+q)^2},
\end{equation}
and solving it for the time derivative of binary separation $\dot{a}$ yields 
\begin{equation}
    \dot{a} = \frac{2(1+q)^2}{q} \sqrt{ \frac{a}{G{M_{\rm b}}^3} } \dot{L}_{\rm b} - \frac{3a}{M_{\rm b}} \dot{M}_{\rm b} - \frac{2a (1-q)}{q(1+q)} \dot{q}.   
\label{adot}
\end{equation}
According to 
\begin{equation}
\dot{a} \propto -\dot{q}
\end{equation}
a drastic decrease in mass ratio corresponds to a shrinking binary separation. Together with the effect of mass accretion 
\begin{equation}
\dot{a} \propto -\dot{M},
\end{equation}
the binary separation is expected to shrink most significantly briefly after the formation of the companion before the change in separation becomes milder, when $\dot{q}$ and $\dot{M}$ decrease.

In fact, our results are in good agreement with results from 3D MHD simulations using \flash\ \citep{Fryxell2000} explicitly considering the protostellar regime \citep{Kuruwita2017}. Starting from idealized spherical cloud conditions \cite{Kuruwita2017} find a quick decrease in binary separation during the early accretion phase of the binary similar to our results.


\subsection{Limitations of the model}

\textbf{Single model run:}
Considering the evolution of the binary/triple system, our results shows a sequence of protostellar multiples involving turbulent fragmentation and protostellar migration. However, we only analyzed a single system with a modest resolution using a minimum cell size of initially $\Delta x \approx 2$ au, and mostly $\Delta x \approx 2$ au for the densest gas. Carrying out comparison runs with a broad appliance of higher resolution for a longer time than to only test the formation of the companions is computationally too expensive currently. 

\textbf{Outflows and sink implementation:}
Outflows are driven mostly on scales of $1$ to 10 au \citep{Bacciotti2002,Bjerkeli2016}, and we, at best, barely resolve mass loss via jets or winds and lack the corresponding feedback \citep{Wang2010,Cunningham2018}. Nevertheless, to account for the mass loss, we simply reduce the mass that accretes onto the sinks by a factor of 2. Given that the evolution of a multiple system depends on the mass accretion rate as well as on the mass ratios of the different components, a thorough analysis of the early evolution of multiple stellar systems requires higher resolution as well as a careful treatment of the accretion onto the sink. For an analysis and discussion of the sink settings and their effect on the formation of stellar multiples, please refer to \cite{Haugboelle2018}. Furthermore, the dynamics of multiples with a separation of $\lesssim 100$ au are also affected by the individual disks of the different components. With our current resolution, we can only roughly account for disks. Finally, our results based on one multiple system can only be suggestive. Constraining the distribution of protostellar separation in detail requires a larger sample of objects.   

\textbf{Magnetic fields and non-ideal MHD effects:}
Regarding magnetic fields, a short-coming of our simulations is the assumption of ideal MHD, and the corresponding negligence of physical resistivities corresponding to Ohmic dissipation, ambipolar diffusion and the Hall effect, \citep[see e.g.][]{Tomida2015,Tsukamoto2015b,Masson2016,Wurster2018}. 
Similar to previous spherical collapse simulations solving the equations of ideal MHD \citep{Seifried2011,Joos2012}, the pile-up of magnetic pressure during the stellar collapse phase causes outward motions of gas away from the sink. Although these magnetic bubbles can lead to compression of gas around the sinks \citep{Vaytet2018}, 
the formation of the arcs -- and eventually the companions -- are ultimately caused by the turbulent dynamics present in the protostellar environment. 
Nevertheless, we aim to avoid potentially spurious effects induced by the magnetic interchange instability by accounting for non-ideal MHD effects in future simulations with the code framework \dispatch\, \citep{Nordlund2018}.

\textbf{Radiative transfer:}
In our model, we model the thermodynamics with a heating and cooling table though the recipe typically causes quasi-isothermal conditions ($T \approx 10$ K) for the densest gas responsible for star formation.
A more sophisticated treatment of the thermodynamics would provide additional thermal support against fragmentation. 
First, the compression of gas itself induces some heating that we most likely underestimate.
However, considering that the collapse phase finalizing companion formation occurs on spatial scales of only a few $10^2$ au (cf. \Fig{comp1+2_slc}), we doubt that additional thermal pressure support would counteract the compression acting on the large-scale sufficiently.  
Second, accounting for the irradiation from nearby stars, \citep[e.g.][]{Geen2015}, in particular the primary star by using a radiative transfer implementation \citep{Rosdahl2013,Rosdahl2015,Frostholm2018} would heat up the gas in the region around the protostar. 

Considering an optically thin environment, the temperature induced by the protostar irradiating as a perfect black body follows 
\begin{equation}
T(r) = \left( \frac{L}{16 \pi \sigma_{\rm SB} r^2} \right)^{1/4},
\label{Tr}
\end{equation}
where $L$ is the luminosity, $\sigma_{\rm SB}$ is the Stefan-Boltzmann constant and $r$ is the radial distance from the protostar. 
Hence, the temperature would drop with increasing radial distance from the star as $T \propto r^{-0.5}$. 
The luminosity of a protostar in its early stage is predominatly determined by the accretion rate. With an accretion rate of $\dot{M} = 10^{-5}$ M$_{\odot}$ yr$^{-1}$ and given a commonly assumed protostellar radius of $R=3$ R$_{\odot}$ \citep{Stahler1988} with mass $M=0.5 M_{\odot}$ its luminosity according to 
\begin{equation}
    L_{\rm acc} = \frac{G M \dot{M}}{R}
\end{equation}
is $L_{\rm acc} \approx 50$ L$_{\odot}$. This rough approximation shows that even for the highest accretion rates, when the primary has a mass of $M\approx 0.5$ M$_{\odot}$, protostellar heating would only modestly increase the temperature beyond $1000$ au distances to less than 30 K. For future studies investigating the processes on smaller distances $r < 10$ au, however, protostellar heating, and radiative transfer are essential.   

\subsection{Comparison with observed arcs and bridges}
In our model, the most outstanding bridge-structure (see upper panel \Fig{prj_bridge2}) occurs a few kyr before the formation of the third companion. The arc connects companion B -- by that time only $\sim$100 au away from the primary -- with the blob that leads to the formation of companion C at a projected distance of about 2000 au. 
The arc resembles a bended bridge with kinematically mild velocity structure compared to its surroundings. 
However, we also see another shorter lived $\lesssim 1000$ au kinematically quiescent bridge-like structure at earlier times ($t\approx t_{\rm B,0} + 10$ kyr) connecting the primary A with the secondary B \Fig{prj_bridge1}.  
The synthetic bridge shown in \Fig{prj_bridge1} is about 1000 au in length. The synthetic bridge shown in \Fig{prj_bridge2} extents to about 2000 au and involves altogether three protostellar sources. Our modeled structure shows several features that are in good agreement with observed arc- and bridge-like structures.

Bridge or arc-like structures have been reported for several deeply embedded sources.
One of the most prominent examples of an observed bridge-like structure is the case of the young binary IRAS 16293, where the two sources are connected by an arc-like filament. The two protostars have a projected distance of 705 au indicating formation via turbulent fragmentation such as is seen in our models. 
The bridge in IRAS 16293 is kinematically quiescent, while the surrounding is kinematically active \citep{Oya2018,vanderWiel2019} similar to the bridge-like structure in our model of the forming triple system. 
Another arc structure is seen for the Class I system IRAS 04191+1523, where a bridge connects the two binary components (projected separation of 860 au). Using C$^{18}$O as a kinematic tracer, \cite{Lee2017} also favor a scenario where the system formed via turbulent fragmentation. 

Different from our triple system, IRAS 04191+1523 consists of only two protostellar components. However, bridge- and arc-structures are also observed for protostellar multiples of higher order than binary. For IRAS 16293, it is debated whether source A is in fact a single protostar \citep{Wootten1989,Chandler2005}, a tight binary \citep{Loinard2007,Pech2010}, or even a tight triple system \citep{Hernandez-Gomez2019} with strong jet components \citep{Kristensen2013,Girart2014,Yeh2008}. A confirmed triple system is the case of SR24 \citep{Fernandez-Lopez2017}. SR24 consists of a close binary SR24N with a separation of only $\sim$10 au \cite{Correia2006} and a third component with a separation of more than 620 au. 
Another striking example of a bent filamentary arm in a triple system is the case of L1448 IRS3B \citep{Tobin2016Natur}. 
With projected separations from the primary of the first companion of 61 au and 183 au of the second companion, the system is more compact than our model as well as the binary systems IRAS 16293 and IRAS 04191+1523. \cite{Tobin2016Natur} show that this system may have been a result of fragmentation on disk rather than turbulent fragmentation on larger scales. 
However, protostellar companions may subsequently migrate and the velocity profile of a multiple system continuously becomes more Keplerian during the capturing phase \citep{Bate2002}. Therefore, L1448 IRS 3B and SR24 -- even involving its close binary SR24N -- may nevertheless have formed via turbulent fragmentation in a similar manner as the wide triple system in our case study.

While most of the observations mentioned above show evidence of bridge structures connecting already formed protostars, the bridge in our model already exists, and is in fact most outstanding prior to the formation of the third companion. This is consistent with observations of prominent arc-structures observed for other embedded sources. 

The two components of IRAS 16293 have been shown to have differences which could be attributed to differences in age. The lack of outflows from source B and prominent outflows observed from source A, have been suggested as a sign that source A is the more evolved object (e.g., \citealt{Pineda2012,Loinard2013,Kristensen2013}). Other indicators, such as chemical differentiation between the sources could also be attributed to age differences, although these differences would indicate source B to be the more evolved source (see \citealt{Calcutt2018a} and \citealt{Calcutt2018b} for a discussion). 
    
Tracing HCO$^+$, \cite{Tokuda2014} observed an arc-structure for L1521F extending from source MMS1 to a distance of $\sim$2000 au with features of small dense cores located in the arc. 
Considering that the second synthetic bridge-structure is most pronounced before the small core has collapsed to form companion C, we expect dispersal of the arc seen in L1521F over the next few $\sim$10 kyr.
\cite{Pineda2015} demonstrate the presence of filamentary structures on scales of $\sim$1000 au around at least one embedded protostar located in the dense core Barnard 5. Their observations show the presence of three density enhancements in these filamentary arms. 
Given the abundance of filamentary structures accompanying star formation in our model, our results support the interpretation that these density enhancements are associated with prestellar condensates.   

Recently, \cite{Sadavoy2018} measured dust polarization in IRAS 16293 to study the morphology of the underlying magnetic field. Analyzing the magnetic field structure in our synthetic bridges is of high interest, but beyond the scope of this paper. Dust polarization depends on the active mechanism of aligning the dust grains, which is rather complex to model in such a dense and turbulent environment. Therefore, instead of providing an oversimplified polarization map based on the magnetic field structure, we present the work of careful synthetic dust polarization measurements with the radiative transfer tool \polaris\ \citep{Reissl2016} in an upcoming paper.

Taking into account all of the above observations, a picture emerges, in which arcs and bridges occur at different evolutionary stages of the formation of protostellar multiples. The temporary appearance of arc- and bridge-structures in our model are consistent with the observations. Our zoom-in model demonstrates that kinematically quiescent bridge-structures are transient phenomena induced by the turbulent motions involved in the formation process of stellar multiples. Our analysis suggests lifetimes of the observed structures of the order of up to a few $10^4$ yr. Although this may seem to be rather short, our model suggests that these structures are common features of the formation of stellar multiples. Therefore, we expect to observe more bridge-like structures around other Class 0 objects considering a duration of the Class 0 phase of approximately $10^5$ yr, and considering that multiple bridge structures can occur during the formation of a protostellar multiple as shown in this paper. Considering that $>50 \%$ of Class 0 systems appear to be multiples \citep{Tobin2016_multisurvey}, together with lifetimes of the Class 0 phase of $\approx 10^5$ yr and the lifetime of the bridges of $\sim10^4$ yr, we expect to see bridge-like structure in $>$ 5\% of Class 0 protostars.

\section{Conclusion}
Using zoom-in simulations, we analyze the formation process of a triple-star system embedded in the turbulent environment of a magnetized GMC. The first companion B forms at $t\approx 35$ kyr after the primary A at a distance of about $1500$ au from the primary and the tertiary C forms at a distance of about $2100$ au from the, by then, more narrow binary system ($r_{\rm AB} \sim$100 au) about 75 kyr after primary A formed. 
Both companions form as a consequence of compression induced by colliding flows associated with turbulent fragmentation in the interstellar medium. 
Our model shows the following sequence for the formation of protostellar multiples: the protostellar companions initially form with a wide separation from the primary ($\sim$1000 au) via turbulent fragmentation, afterwards migrate inwards to distances of $\sim$100 au on timescales of $\Delta t \sim 10$ kyr before they are captured and bound in eccentric systems of protostellar multiples.
Once the system is bound, the accretion profiles of the young protostars are variable related to the periodic pattern of the orbital frequency of the system.  

We find transient filamentary arms connecting two protostars that build as a by-product of the formation process of the companions. These bridges persist for time-scales of the order of $\Delta t\sim 10$ kyr. 
Studying the properties of these `bridges' more closely shows no sign of a preferred motion toward any of the protostellar components.
Instead, the velocity components of the colliding flows cancel out and the bridge becomes kinematically quiescent, similar to what has recently been observed in systems such as IRAS 16293--2422 \citep{vanderWiel2019}. 

Considering the velocity components, our analysis shows that bridge-structures are a consequence of compression due to flows acting on larger scales, partly cancelling out the velocity components in the compressed region forming the bridge. In this way, the gas located inside the bridge can become kinematically rather quiescent compared to the systemic velocity.  
With respect to the accretion process of the companions, the bridge structure acts as an important important mass reservoir of the different stellar components. 
Using tracer particles, we analyze the origin of the gas accreting onto the different components. The analysis shows: 
\begin{itemize}
    \item the different protostellar components -- at least partly -- share the same mass reservoir, and
    \item the protostellar companions are fed by the gas located in the elongated compressed filament.
\end{itemize} 
Therefore, the gas located in the bridge eventually contributes to protostellar accretion in the system, but it is different from a gas stream feeding one individual source.
While the gas in streams actively approaches a single star from one direction, the gas located in the bridge is available to be picked up by any star in the system. Gas located in different parts of the bridge can accrete onto one of the sources, and hence the bridge may consist of gas streams with flow directions toward different sources.

In this paper, we aimed for a deeper understanding of the origin of arc- and bridge-like structures observed for multiple embedded protostars.
In particular, the origin of quiescent dense structures (e.g., IRAS 16293-2422) is difficult to understand with a picture of isolated star formation in mind. 
However, accounting for the overall dynamics in the turbulent GMC, the results bring to light that such structures are induced by the underlying turbulent motions in the GMC. Our model demonstrates that bridge-like structures occur as natural transient phenomena associated with the formation of protostellar multiples via turbulent fragmentation.
Against the background of observed arc- and bridge-like structures associated with protostellar multiples, our results strongly indicate age differences of $\Delta t \sim 10$ kyr between the different components of the multiple. Future kinematic studies of young protostars in bridge structures will help to test this result. 


\begin{acknowledgements}
We thank the anonymous referee for their insightful and constructive comments on an earlier draft of the manuscript that helped to improve the quality of this paper.
We thank Troels Haugb{\o}lle and {\AA}ke Nordlund for their development of the zoom-in technique for the modified version of \ramses. 
MK thanks the developers of the python-based analyzing tool YT http://yt-project.org/ \citep{yt-reference}, which simplified the analysis significantly.
Furthermore, MK thanks Kees Dullemond for insightful discussions. 
The research of MK is supported by a research grant of the Independent Research Foundation Denmark (IRFD) (international postdoctoral fellow, project numer: 8028-00025B). We acknowledge PRACE for awarding access to the computing resource CURIE based in France at CEA for carrying out part of the simulations. Thanks to a research grant from Villum Fonden (VKR023406), MK could use archival storage and computing nodes at the University of Copenhagen HPC centre to carry out essential parts of the simulations and the post-processing.
MK acknowledges the support of the DFG Research Unit `Transition Disks' (FOR 2634/1, ER 685/8-1). The research of LEK is supported by a research grant (19127) from VILLUM FONDEN. Research at the Centre for Star and Planet Formation is funded by the Danish National Research Foundation.
\end{acknowledgements}



\begin{appendix}
\section{Formation of companions at higher resolution}
\begin{figure}
  \centering
  \includegraphics[width=0.5\textwidth]{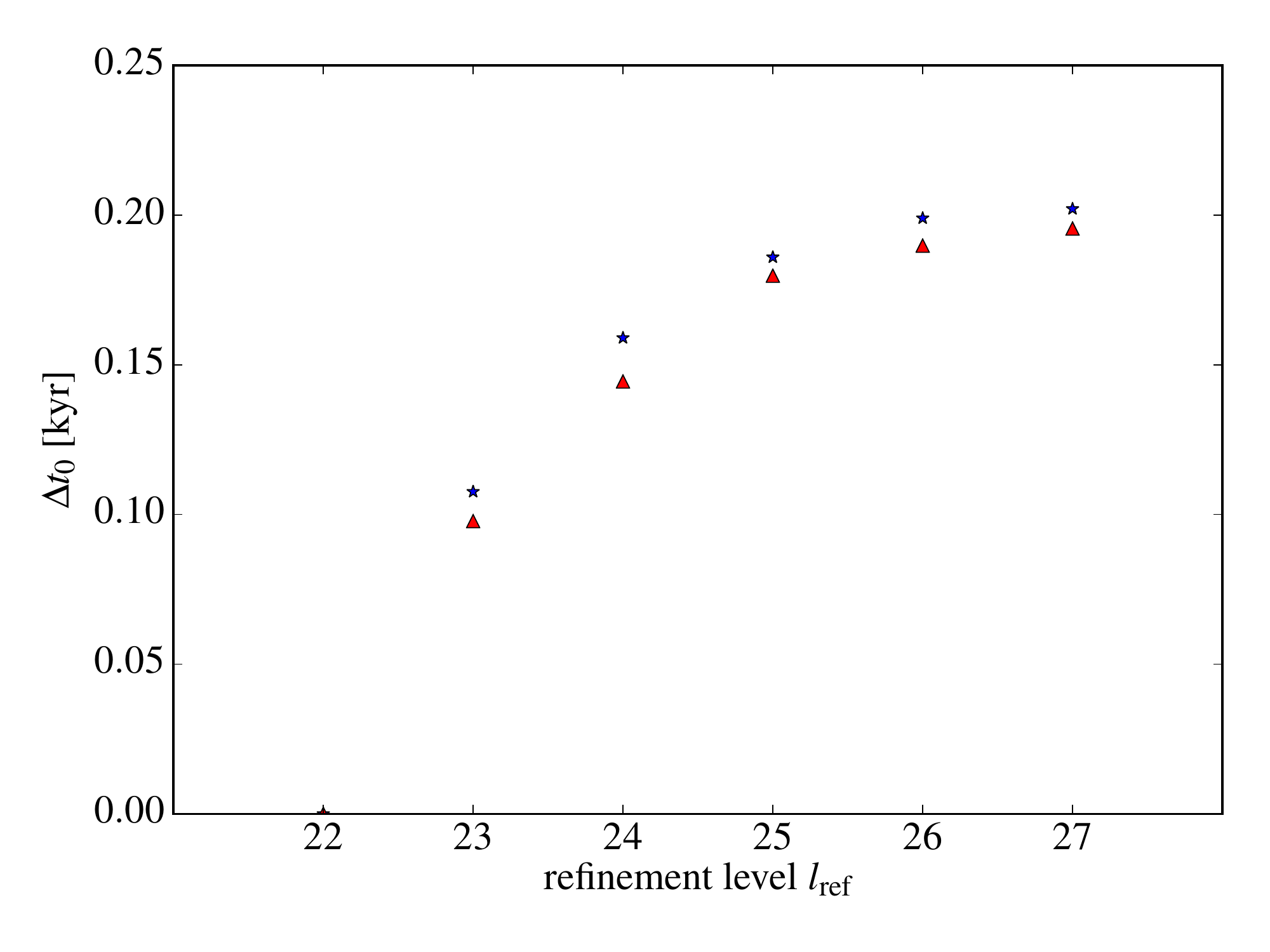}
  \caption{The plot shows the time of formation of sinks A (blue asterisks) and sink B (red triangles) using different maximum resolution with respect to sink formation using a resolution of $l_{\rm ref}=22$ corresponding to a minimum cell size of $2$ au. }
  \label{fig:s_t0}
\end{figure}

Stars form as a consequence of gravitational collapse. In our numerical scheme, we account for these properties by requiring gas to be above a given density threshold as well as the gas in the cell, i.e., infalling gas $\nabla \cdot \mathbf{v} < 0$.
In a dense turbulent medium, using low resolution averages out the deviations of the velocity field. 
As a consequence, sinks that form at low resolution, may not form at higher resolution when accounting for the velocity deviations.
As mentioned in the text, the system forms in a turbulent medium with fluctuating velocities.  
To test, whether the formation of the companions is robust, we conducted comparison runs with higher resolution in the regions, where companion B or C form. For the test, we use $l_{\rm ref}=22$(23,24,25,26,27) corresponding to minimum cell sizes of $\approx2$ au ($\approx1$ au, $\approx0.5$ au, $\approx0.23$ au, $\approx0.123$ au, $\approx0.061$ au). As shown in \Figure{s_t0}, sink formation is triggered in the higher resolution runs demonstrating that the sinks indeed form due to a local collapse on smaller scales triggered by the colliding flows acting on larger scales. Sinks form later when using higher resolution because the density to trigger sink formation is a multiple  of the cell density at highest level. The density threshold for triggering the formation of a sink is $10\times \rho_{\rm c}$, where $\rho_{\rm c}$ is the density threshold for resolving a cell to highest resolution. To form a sink, the threshold density has to be refined with at least 25 cells. As the densities increase during protostellar collapse with evolving time, applying higher resolution therefore delays the creation of the sink particle. However, for the refinement levels considered here, the delay is $<1$ kyr, and hence negligible for our analysis of the evolution on time scales of up to 100 kyr \citep[see also][]{Kuffmeier2017}. 

\end{appendix}
\end{document}